\documentclass[11pt,reqno,oneside]{amsart}
\usepackage{amssymb}
\usepackage[reqno]{amsmath}

\input{doublespace.sty}
\begin{document}

\title{The Tetrad Frame Constraint Algebra}
\author{M.A. Clayton\footnote{Present address: CERN--Th, CH--1211 Geneva 23, 
Switzerland}}
\address{Department of Physics, University of Toronto, Toronto, \textsc{on}, 
Canada M5S 1A7}
\email{clayton@medb.physics.utoronto.ca}
\date{\today}
\thanks{UTPT-96-15}

\begin{abstract}
It is shown via the principle of path independence that the (time gauge) 
constraint algebra derived in \cite{Charap+Henneaux+Nelson:1988} for vielbein 
General Relativity is a generic feature of any covariant theory formulated in 
a vielbein frame.
In the process of doing so, the relationship between the coordinate and 
orthonormal frame algebera is made explicit.
\end{abstract}
\maketitle

\section*{Introduction}
\label{sect:Intro}

It is by now well-known that the canonical constraint algebra for any 
covariantly constructed field theory is of a fixed form regardless of the 
theory in question~\cite{Teitelboim:1973}.
This algebra reflects the embeddability of a spatial surface into the 
spacetime manifold, that is, whether the theory is consistent with the 
foliation of spacetime into arbitrary spacelike hypersurfaces; an assumption 
that has been used in a derivation of 
Geometrodynamics~\cite{Hojman+Kuchar+Teitelboim:1976}.
These results were derived using coordinate frames on the spatial 
hypersurface, and the original translation of the algebra to an orthonormal 
frame in tetrad gravity~\cite{Nelson+Teitelboim:1978} was not complete.
The correct algebra~\cite{Henneaux:1983,Charap+Henneaux+Nelson:1988} was 
derived within a specific model and therefore not obviously a generic 
feature of tetrad theories. 
Here it will be shown that the (time gauge) constraint algebra given 
in~\cite{Charap+Henneaux+Nelson:1988} may be derived using similar arguments 
to those in~\cite{Teitelboim:1973}, and is therefore generic.
Since the result must hold on the entire phase space (not just on the 
constraint surface), it is important for any quantisation that considers 
states that do not satisfy the constraints~\cite{Henneaux:1985}.

To prove this result we will proceed in three stages, each with an associated 
section.
First the standard geometry of embeddings is reviewed and extended to the 
case where one is considering an arbitrary (nondegenerate) linear frame on 
the spatial hypersurface.  
We also give the description of the hypersurface geometry that we will be 
using throughout.
Next we review and generalise the principle of path independence that relates 
the diffeomorphism constraint algebra to the geometry of the hypersurface.
The bulk of the paper is devoted to the third section, in which we consider 
the generators of hypersurface deformation as a means of determining the 
structure functions that appear in the constraint algebra. 
In doing so, we consider two situations explicitly: the first where the frame 
is chosen to be independent of the embedding, yeilding results that may be 
easily related to the original coordinate frame approach (which is a special 
case), and the second where the form of the metric is left unaltered by the 
action of the deformation generators, from which a limit to an orthonormal 
frame is straightforward.

\section{the Embeddings and Hypersurface Geometry}
\label{sect:Vielbein}

The basic object throughout this work will be the embedding 
$\mathsf{e}\in\mathit{Emb}_\mathrm{g}(\Sigma,\mathbf{M}):
\Sigma\rightarrow\mathbf{M}$, which maps an $n$-dimensional, Riemannian 
manifold $\Sigma$ into an $n+1$-dimensional, pseudo--Riemannian (Lorentzian) 
manifold $\mathbf{M}$.
In coordinate charts represented by the coordinates $x^i$ on $\Sigma$ and 
$x^\mu$ on $\mathbf{M}$ (lowercase Greek and Roman indices will represent 
spacetime and spatial coordinate indices throughout), the embedding takes 
the form $\mathsf{e}:x^i\rightarrow \mathsf{e}^\alpha(x^i)$.
Although we will not discuss the geometry of hyperspace explicitly herein, 
much of the formalism is directly adapted from 
Kucha\v{r}~\cite{Kuchar:1976a,Kuchar:1976b} with slight changes in notation.

We introduce an arbitrary linear frame (and dual coframe) on $\Sigma$ related 
through the vielbein ${E_a}^i$ to a coordinate frame and coframe 
as~\cite{Nakahara:1990} $E_{ax}:={E_a}^i(x)\partial_{x^i}$ and 
$\theta^a(x):=dx^i{E_i}^a(x)$ respectively, where the vielbeins satisfy the 
frame duality conditions ${E_b}^i{E_i}^a=\delta^a_b$ and 
${E_j}^a{E_a}^i=\delta^i_j$, and $\partial_{x^i}$ indicates the partial 
derivative with respect to $x^i$ on $\Sigma$.
(Note that by construction we are singling-out a normal frame vector, and so 
the results derived will correspond to the time gauge 
of~\cite{Charap+Henneaux+Nelson:1988}.)
We will use $a,b,c\ldots\in\{1,2,3,\ldots n\}$ to indicate the components of 
a tensor in this surface frame (all results herein will apply to 
hypersurfaces of arbitrary dimension $n$), and 
$A,B,C\ldots\in\{0,1,2,3,\ldots n\}$ for a frame above $\mathbf{M}$.
This linear frame and its dual are used as a basis in the tangent bundle 
$T\Sigma$ and tensor bundles associated to it.
This is by now a standard 
procedure~\cite{Kobayashi+Nomizu:1963,Choquet-Bruhat+:1989} which has been 
applied to general relativity~\cite{Komar:1984,Floreanini+Percacci:1990}. 

The transformations that (locally) make a change of 
frame---or frame rotation---on $\Sigma$ are elements of 
$\mathrm{GL}(n,\mathbb{R})$, acting on the frame and coframe (and similarly 
on the components of tensors) as: $e_a\rightarrow\lvert M^{-1}\rvert^b_ae_b$ 
and $\theta^a\rightarrow M^a_b\theta^b$, where $M\in\mathrm{GL}(n,\mathbb{R})$.
Considering the infinitesimal form of these transformations 
$M^a_b\rightarrow\delta^a_b+\Omega^a_b$, one finds the the Lie algebra 
$\mathfrak{gl}(n,\mathbb{R})$ is generated by an operator $\Delta^a_{bx}$, 
defined to act on vectors and covectors as (extendible to arbitrary tensors 
in the usual manner)
\begin{equation}\label{eq:rotat}
\Delta^a_{bx}[V_c(y)]=-\delta^a_cV_b(y)\delta(y,x),\quad
\Delta^a_{bx}[V^c(y)]=\delta^c_bV^a(y)\delta(y,x),
\end{equation}
and thus $\Omega^a_b(x)\Delta^b_{ax}$ is the infinitesimal form of the frame 
rotation considered above.
The Lie algebra of these frame rotation generators is straightforward to 
compute, yielding the standard result
\begin{equation}\label{eq:Gl3}
[\Delta^a_{bx},\Delta^c_{dy}]
=\delta^c_b\Delta^a_{dx}\delta(x,y)-
\delta^a_d\Delta^c_{by}\delta(y,x).
\end{equation} 
Note that throughout we will be considering the full set of 
$\mathfrak{gl}(n,\mathbb{R})$ generators; reducing the frame bundle to 
$\mathrm{SO}(n)$ and considering therefore the generators of $n$-dimensional 
rotations is equivalent to considering only the antisymmetric generators.

Since the embedding clearly maps curves in $\Sigma$ to those in $\mathbf{M}$, 
it induces the pullback map 
$\mathsf{e}_*:T_{\mathsf{e}(x)}\mathbf{M}\rightarrow T_x\Sigma$ which has the 
local form
\begin{equation}\label{eq:emb pf}
\mathsf{e}^\alpha_a(x)
=E_{ax}[\mathsf{e}^\alpha(x)]
={E_a}^i(x)\partial_{x^i}[\mathsf{e}^\alpha(x)],
\end{equation}
where the presence of the vielbein is required in order to map the components 
of vectors in $T_{\mathsf{e}(x)}\mathbf{M}$ to vectors in $T\Sigma$ that are 
expanded in the chosen linear frame.
The set of vectors $\mathsf{e}_a:=\mathsf{e}^\alpha_a\partial_{x^\alpha}$ 
defines an $n$-dimensional subspace of $T\mathbf{M}$, the remaining dimension 
spanned by the normal vector~\cite{Isenberg+Nester:1980}, defined to be a 
unit vector in $T_{e(x)}\mathbf{M}$ that is orthogonal to $\mathsf{e}_a$:
\begin{equation}\label{eq:normal}
\mathrm{g}(\mathsf{n},\mathsf{e}_a)=
\mathsf{n}_\alpha\mathsf{e}^\alpha_a=0,\quad 
\mathrm{g}(\mathsf{n},\mathsf{n})=
\mathsf{n}^\alpha\mathsf{n}_\alpha=1.
\end{equation}
These conditions combined with the fact that the spacetime metric 
$\mathrm{g}_{\alpha\beta}$ has been chosen to have $(+,-,-,-,\ldots)$ 
signature, reflect the fact that $\Sigma$ is a spacelike embedded surface.
The chosen normal vector combined with the pullback map (viewed as $n$ 
vector fields on $\mathbf{M}$) $\{e_A\}:=\{\mathsf{n},\mathsf{e}_a\}$ define 
a frame on which all spacetime tensors may be decomposed, separating them 
into into normal and tangential components to $\Sigma$ respectively.
For example, a vector field is decomposed as 
$V^\alpha=V_{\mathsf{n}}\mathsf{n}^\alpha+\mathsf{e}^\alpha_aV^a$, where 
$V_{\mathsf{n}}:=V^\alpha\mathsf{n}_\alpha$ is the scalar component 
perpindicular to $\Sigma$ and $V^a:=V^\alpha\mathsf{e}_\alpha^a$ are the 
vector components that are tangential to $\Sigma$.
The (negative-definite) metric over $\Sigma$ is defined as the pullback of 
the spacetime metric 
$\mathrm{g}_{ab}:=[\mathsf{e}_*\mathrm{g}]_{ab}=
\mathrm{g}_{\alpha\beta}\mathsf{e}^\alpha_a\mathsf{e}^\beta_b$.
The spacetime metric takes on the projected form 
$\mathrm{g}_{\alpha\beta}=\mathsf{n}_\alpha\mathsf{n}_\beta
+\mathrm{g}_{ab}\mathsf{e}^a_\alpha\mathsf{e}^b_\beta$, or in terms of the 
dual basis $\{\mathsf{n},\mathsf{\theta}^a\}$ 
(where $\mathsf{\theta}^a[\mathsf{e}_b]=\delta^a_b$) is 
$\mathrm{g}=\mathsf{n}\otimes\mathsf{n}+\mathrm{g}_{ab}\mathsf{\theta}^a
\otimes\mathsf{\theta}^b$.
Throughout Greek and Roman indices are `raised' and `lowered' using 
$\mathrm{g}_{\alpha\beta}$ and $\mathrm{g}_{ab}$ respectively.
This projection is extendible to higher-order tensors in a straightforward 
manner~\cite{Kuchar:1976a}.

The foliation of $\mathbf{M}$ by non-overlapping spacelike hypersurfaces is 
accomplished by introducing a family of embeddings $\mathsf{e}(t)$ that 
cover $\mathbf{M}$ indexed by a coordinate $t$, effectively realizing the 
diffeomorphism $\mathsf{e}(t,x):\mathbb{R}\times\Sigma\rightarrow \mathbf{M}$.
The resulting family of pullback maps then give 
$\mathsf{e}_*(t,x): T_{\mathsf{e}(t,x)}\mathbf{M}\rightarrow 
T_t\mathbb{R}\times T_x\Sigma$, where a tangent to $\mathbb{R}$ is related to 
the lapse function $\mathsf{N}$ and shift vector $\mathsf{N}^a$ by
\begin{equation}\label{eq:Ns}
\mathsf{N}:=\mathsf{n}_\alpha\partial_t[\mathsf{e}^\alpha],\quad
\mathsf{N}^a:=\mathsf{e}_\alpha^a\partial_t[\mathsf{e}^\alpha].
\end{equation}
This decomposition makes clear that the lapse function and shift vector 
represent the normal and tangential (to $\Sigma$) respectively of the time 
vector $\partial_t$.
It is also clear that the local `flow of time' (or choice of family of 
embeddings) near a hypersurface is parameterised by a choice of $\mathsf{N}$ 
and $\mathsf{N}^a$.
A sans-serif font will  be used in order to indicate nontrivial dependence 
on the embedding (\textit{i.e.}, not just via the surface projection); 
notable exceptions to this will be the extrinsic curvature $k_{ab}$ and in
trinsic connection coefficients $\Gamma^a_{bc}$.

The coordinate frame inverse of the pull-back map: 
$\frac{\partial t}{\partial x^\alpha}=\frac{\mathsf{n}_\alpha}{\mathsf{N}}$ 
and $\frac{\partial y^i}{\partial x^\alpha}=\mathsf{e}^i_\alpha
-\frac{\mathsf{N}^i\mathsf{n}_\alpha}{\mathsf{N}}$, may be used to make the 
mapping between $\{\partial_t,E_a\}$ and the spacetime, surface-compatible 
frame $\{\mathsf{n},\mathsf{e}_a\}$:
\begin{equation}\label{eq:translate}
\mathsf{n}=\partial_{\mathsf{n}}:=\mathsf{n}^\alpha\partial_\alpha
=\mathsf{N}^{-1}(\partial_t-\mathsf{N}^aE_a),\quad
\mathsf{e}^\alpha_a\partial_\alpha
=E_a={E_a}^i\partial_{i},
\end{equation}
may be derived.
This allows us to do a projection of the spacetime geometry onto 
$\Sigma$~\cite{Isenberg+Nester:1980} and in general do away with the 
spacetime coordinates $x^\alpha$ altogether, so that, for example, the 
argument of $F(x)$ indicates dependence on on a point in the spatial 
hypersurface.

In the frame $\{\mathsf{n},\mathsf{e}_a\}$ we may compute the Levi-Civita 
connection coefficients from the metric compatibility conditions and the 
condition of vanishing torsion, to find: 
$\Gamma^\mathsf{n}_{\mathsf{n}\mathsf{n}}=\Gamma^\mathsf{n}_{a\mathsf{n}}=0$, 
$\Gamma^\mathsf{n}_{\mathsf{n}a}:=a_a
=-\mathrm{g}_{ab}\Gamma^b_{\mathsf{n}\mathsf{n}}$, 
$\Gamma^\mathsf{n}_{ab}=\Gamma^\mathsf{n}_{ba}:=k_{ab}
=-\mathrm{g}_{ac}\Gamma^c_{b\mathsf{n}}$, and 
$\Gamma^a_{b\mathsf{n}}=\Gamma^a_{\mathsf{n}b}+{C_{b\mathsf{n}}}^a$.
In these, we have used~\eqref{eq:translate} to derive the nonvanishing 
structure constants of the pulled-back basis vectors: 
$[\mathsf{e}_a,\mathsf{e}_b]={C_{ab}}^c\mathsf{e}_c$ and 
$[\mathsf{n},\mathsf{e}_a]=a_a\mathsf{n}+{C_{\mathsf{n}a}}^b\mathsf{e}_b$, 
where ${C_{ab}}^c={E_i}^c(E_a[{E_b}^i]-E_b[{E_a}^i])$ are the structure 
constants for the basis defined on $\Sigma$, and 
\begin{subequations}\label{eq:a C}
\begin{gather}
a_a:={C_{\mathsf{n}a}}^{\mathsf{n}}
=-\Gamma^\nu_{\mu\beta}\mathsf{n}^\mu\mathsf{n}^\beta\mathsf{e}_{\nu a}
=\mathsf{n}_\mu\partial_{\mathsf{n}}[\mathsf{e}^\mu_a]
-\mathsf{n}_\mu E_a[\mathsf{n}^\mu]=E_a[\ln\mathsf{N}],\\
{C_{\mathsf{n}a}}^b=
\mathsf{e}^b_\mu\partial_{\mathsf{n}}[\mathsf{e}^\mu_a]
-\mathsf{e}^b_\mu E_a[\mathsf{n}^\mu]
=\mathsf{N}^{-1}\bigl(E_a[\mathsf{N}^b]+\mathsf{N}^c{C_{ac}}^b
+\partial_t[{E_a}^i]{E_i}^b\bigr).
\end{gather}
\end{subequations}
All of these quantities have been written in terms of the hypersurface 
geometry and the local parameterisation of the family of embeddings via 
$\mathsf{N}$ and $\mathsf{N}^a$.

In (\ref{eq:a C}), we have introduced the coordinate components of the 
Levi-Civita connection on $\mathbf{M}$, which may be related to that on 
$\Sigma$ by 
\begin{equation}\label{eq:Gammer}
\Gamma^a_{bc}=\mathsf{e}^a_\mu E_b[\mathsf{e}_c^\mu]
+\mathsf{e}^\mu_b\mathsf{e}^\nu_c\mathsf{e}_\alpha^a\Gamma^\alpha_{\mu\nu}, 
\end{equation}
which satisfies the intrinsic metric-compatibility conditions 
${}^\Sigma\nabla_c[\mathrm{g}]_{ab}:=E_c[\mathrm{g}_{bc}]
-\Gamma^d_{ca}\mathrm{g}_{db}-\Gamma^d_{cb}\mathrm{g}_{ad}=0$. 
The presence of the spacetime metric implies that such a connection is 
uniquely defined, and so may be introduced regardless of whether or not it 
plays a fundamental role in the theory in question. 
Other projections of the spacetime compatibility conditions result in 
$\mathsf{n}^\mu\partial_\alpha[\mathsf{n}_\mu]
=\mathsf{n}_\mu\mathsf{n}^\nu\Gamma^\mu_{\alpha\nu}$, and define the 
extrinsic curvature $k_{ab}$ of $\Sigma$
\begin{subequations}\label{eq:k definition}
\begin{gather}
\partial_{\mathsf{n}}[\mathrm{g}_{ab}]
-2\mathrm{g}_{c(b}{C_{\mathsf{n}a)}}^c+2k_{ab}=0,\\
k_{ab}:=\mathsf{n}_\alpha E_a[\mathsf{e}^\alpha_b]
+\mathsf{n}_\gamma\mathsf{e}^\alpha_a\mathsf{e}^\beta_b
\Gamma^\gamma_{\alpha\beta}
=-\mathsf{e}_{\beta b}E_a[\mathsf{n}^\beta]
-\mathsf{n}^\gamma\mathsf{e}_{\alpha b}\mathsf{e}^\mu_a
\Gamma^\alpha_{\gamma\mu},
\end{gather}
\end{subequations}
which may be shown to be a hypersurface tensor under frame rotations.
Note that this structure is identical to what appears when introducing a 
surface adapted basis~\cite{Isenberg+Nester:1980}, and as in that reference, 
we will use the minimal number of defined quantities $a_a$, $k_{ab}$, 
$\Gamma^a_{bc}$, and ${C_{\mathsf{n}a}}^b$ throughout.

We will also require some of the projected components of the spacetime 
Riemann tensor defined in general by 
${R^A}_{BCD}=E_C[\Gamma^A_{DB}]-E_D[\Gamma^A_{CB}]
+\Gamma^E_{DB}\Gamma^A_{CE}-\Gamma^E_{CB}\Gamma^A_{DE}
-{C_{CD}}^E\Gamma^A_{EB}$.
The components with one normal projection are related by 
${R^b}_{ca\mathsf{n}}=-\mathrm{g}^{bd}{R^{\mathsf{n}}}_{adc}$, where 
${R^a}_{bc\mathsf{n}}:=\mathsf{e}^a_\mu 
\mathsf{e}^\alpha_b\mathsf{e}^\beta_c \mathsf{n}^\gamma 
{R^\mu}_{\alpha\beta\gamma}$ and 
${R^{\mathsf{n}}}_{abc}:=\mathsf{e}_a^\gamma 
\mathsf{e}^\alpha_b\mathsf{e}^\beta_c \mathsf{n}_\mu 
{R^\mu}_{\gamma\alpha\beta}=\nabla_b[k]_{ca}-\nabla_c[k]_{ba}$, 
in particular implying that
\begin{equation}
\begin{split}
{R^a}_{bc\mathsf{n}}
=&-\nabla_c[k]^a_b-k^a_ba_c-k^a_ca_b+k_{cb}a^a
+E_c[{C_{\mathsf{n}b}}^a]
+{C_{\mathsf{n}b}}^aa_c \\
&-\partial_{\mathsf{n}}[\Gamma^a_{cb}]
+{C_{\mathsf{n}b}}^e\Gamma^a_{ce}
-{C_{\mathsf{n}e}}^a\Gamma^e_{cb}
+{C_{\mathsf{n}c}}^e\Gamma^a_{eb} \\
=&\nabla_c[k]^b_a-\mathrm{g}^{bd}\nabla_d[k]_{ca}.
\end{split}
\end{equation}
The intrinsic Riemannian curvature of $\Sigma$ defined in terms of the 
intrinsic connection components $\Gamma^a_{bc}$ will be denoted 
${}^\Sigma\!{R^a}_{bcd}$, and should not be confused with the spatially 
projected components of the spacetime Riemann tensor 
${R^a}_{bcd}:=\mathsf{e}^a_\alpha\mathsf{e}_b^\beta\mathsf{e}_c^\mu
\mathsf{e}_d^\nu{R^\alpha}_{\beta\mu\nu}$.

Note that the results of this work do not require that one consider a metric 
compatible theory nor even a metric theory.
The reason is that for any such theory, one is still required to define a 
normal vector in order to pass to the Hamiltonian formalism.
This definition requires a symmetric tensor that plays the role of a metric 
in~\eqref{eq:normal}, and therefore also defines the Riemannian geometry 
described in this section; it is this metric that will therefore appear in 
the constraint algebra.
Note also that this construction will \textit{not} tell one anything about 
the consistency of the Cauchy problem (as may be seen by the fact that the 
original investigations of the algebra from this point of view consider both 
Lorentzian and Riemannian signatures for the spacetime metric), merely 
whether the dynamics are consistent with the geometry.

\section{The Principle of Path Independence}

The `principle of path independence' expresses the conviction that a physical 
system set up on an initial Cauchy surface $\Sigma_i$ should evolve to a 
unique state on a later hypersurface $\Sigma_f$ regardless of how one looks 
at the evolution 
in-between~\cite{Teitelboim:1973,Hojman+Kuchar+Teitelboim:1976}.
In order to examine the consequences of this principle, note that the lapse 
function and shift vector $\mathsf{N}^A:=(\mathsf{N},\mathsf{N}^i)$ (as 
defined through the family of embeddings by~\eqref{eq:Ns} or as metric 
components in the usual GR definition~\cite{ADM:1959}) are geometrical 
objects that encode the direction in which the surface is evolving in 
spacetime, or, equivalently, how the surface coordinate labels are evolving 
with time.
If we consider infinitesimal evolution from $\Sigma_i$ along $\mathsf{N}^A_1$ 
followed by evolution along $\mathsf{N}^A_2$, the resulting hypersurface 
$\Sigma_f=\Sigma_{12}$ is in general a different hypersurface than 
$\Sigma_{21}$ which results from evolution along $\mathsf{N}^A_2$ followed 
by $\mathsf{N}^A_1$.
The deformation vector $\mathsf{N}^A$ that evolves $\Sigma_{21}$ to 
$\Sigma_f$ will be related to $\mathsf{N}^A_1$ and $\mathsf{N}^A_2$ in 
general through the structure functions $\kappa^A_{BC}$ defined 
by~\cite{Teitelboim:1973}
\begin{equation}\label{eq:structure functions}
\mathsf{N}^A(x)=\int_\Sigma dy_1\int_\Sigma dy_2\,
\kappa^A_{BC}(x;y_1,y_2)\mathsf{N}^B_1(y_1)\mathsf{N}^C_2(y_2).
\end{equation}
Here we assume for simplicity that $\Sigma_{21}$ is completely to the past 
of $\Sigma_f$ so that we may consider further evolution from $\Sigma_{21}$ to 
$\Sigma_f$.
As we shall see, the structure functions $\kappa^A_{BC}$ are determined 
completely from the hypersurface geometry alone (\textit{i.e.}, they are 
intrinsic) and are the same structure functions that appear in the 
constraint algebra (as we will discuss in the following paragraphs).
It is the computation of these structure functions which makes up the bulk 
of the original coordinate frame work~\cite{Teitelboim:1973} as well as the 
present manuscript.

Translating the principle of path independence to a Hamiltonian system is a 
fairly straightforward procedure; we merely consider the evolution of initial 
data on $\Sigma_i$  to data on $\Sigma_f$ along the same two paths considered 
above.
The real content of the argument rests in recognising the fact that the 
general form of the Hamiltonian for a diffeomorphism invariant system may be 
put into a very specific form.
We will begin by reviewing the coordinate frame argument 
of~\cite{Teitelboim:1973,Hojman+Kuchar+Teitelboim:1976}, and afterwards see 
that the generalisation to an arbitrary choice of linear frame is farily 
straightforward.

It is by now well-known that the lapse function and shift vector will appear 
undifferentiated in the action for \textit{any} covarinatly constructed 
Lagrangian.
That this is so follows from the fact that a covariant action allows the use 
of any family of embeddings that cover $\mathbf{M}$, and therefore the local 
parameterisation of the family (the lapse and shift) should be freely 
specifiable.
Therefore considering the case of gravitational theories for which the 
fields that describe the intrinsic geometry appear in phase space as 
canonical coordinates, treating $\mathsf{N}$ and $\mathsf{N}^i$ as Lagrange 
multipliers the Hamiltonain may be written in the form 
$H=\int_\Sigma dx(\mathsf{N}\mathcal{H}+\mathsf{N}^i\mathcal{H}_i)$ 
(modulo surface terms which will be systematically ignored throughout this 
work), immediately resulting in the Diffeomorphism constraints 
$\mathcal{H}\approx\mathcal{H}_i\approx 0$.
Although we will consider this case exclusively, it is not difficult to 
extend the results to parameterised field theories propagating on a fixed 
spacetime background where where $\mathcal{H}$ and $\mathcal{H}_i$ are the 
energy and momentum respectively of the field, as well as to unparameterised 
fields where there is a slight change in the ensuing constraint 
algebra~\cite{Hojman+Kuchar+Teitelboim:1976}.

The evolution of a scalar functional $F$ on phase space as it evolves from 
an initial hypersurface to a final one related to it via the infinitesimal 
deformation given by the parameters $\mathsf{N}^A:=(\mathsf{N},\mathsf{N}^i)$ 
is determined by
\begin{equation}
\delta F=\dot{F}=
\{F,H\}=\int_\Sigma dx\,\{F,\mathsf{N}^A\mathcal{H}_A\},
\end{equation}
where the standard Poisson brackets~\cite{Henneaux+Teitelboim:1992} have been 
assumed.
Using the path independence argument, the change in $F$ as the system evolves 
from $\Sigma_{21}$ to $\Sigma_f$ is given by
\begin{equation}
\begin{split}
F_{\Sigma_{12}}-F_{\Sigma_{21}}
=&\int_\Sigma dx\,\{F,\mathsf{N}^A\mathcal{H}_A\}\\
=&-\int_\Sigma dy_1\,\int_\Sigma dy_2\,
\Bigl\{F,\{\mathsf{N}_1^A(y_1)\mathcal{H}_A(y_1),
\mathsf{N}_2^A(y_2)\mathcal{H}_A(y_2)\}\Bigr\},
\end{split}
\end{equation}
the first of which is the direct evolution using $\mathsf{N}^A$ as defined 
by~\eqref{eq:structure functions}, and the latter is the difference as given 
by the evolution along the alternating paths (the Jacobi identity has been 
used to write it in this form).

Inserting the structure functions~\eqref{eq:structure functions} and noting 
that the lapse and shift may be pulleld through the Poisson brackets (we are 
assuming that the diffeomorphism constraints are satisfied), we find the 
condition
\begin{equation}\label{eq:basic}
\Bigl\{F,\{\mathcal{H}_B(y_1),\mathcal{H}_B(y_2)\}
+\int_\Sigma dx\,\kappa^A_{BC}(x;y_1,y_2)\mathcal{H}_A(x)\Bigr\}=0.
\end{equation}
Removing the functional $F$ from this condition is 
straightforward~\cite{Teitelboim:1973}.
If the argument of the overall Poisson bracket ($\{F,\cdot\}$) depends in a 
nontrivial way on canonical variables, then $F$ may be chosen as a functional 
that is conjugate to it, and~\eqref{eq:basic} would not be satisfied.
Therefore the argument must be a constant on phase space, and since the 
diffeomorphism constraints are satisfied, this constant factor must be 
identically zero, resulting in the condition
\begin{equation}\label{eq:H alg}
\{\mathcal{H}_B(y_1),\mathcal{H}_B(y_2)\}=
-\int_\Sigma dx\,\kappa^A_{BC}(x;y_1,y_2)\mathcal{H}_A(x),
\end{equation}
which is the relation between dynamics and geometry that we have been seeking.

For the more general case at hand, this argument is actually extended in a 
very straightforward manner.
Since we have introduced an arbitrary frame on $\Sigma$, the Lagrangian will 
posess a $\mathrm{GL}(n,\mathbb{R})$ symmetry in addition to diffeomorphism 
invariance.
Using the argument in~\cite{Floreanini+Percacci:1990} we know that variations 
with respect to the spatial vielbeins ${E_i}^a$ are not independent of those 
with respect to the surface metric $\mathrm{g}_{ab}$, and therefore when 
passing to the Hamiltonian formalism there will be $n^2$ constraints that 
must be imposed via Lagrange multipliers $\mathsf{N}^a_b$ (which may be 
chosen to be the time component of the Ricci coefficient in tetrad 
gravity~\cite{Charap+Henneaux+Nelson:1988}).
Since the lapse and shift play exactly the same role as they did previoulsy, 
the Hamiltonian takes the form
\begin{equation}\label{eq:aitch}
H=\int_\Sigma dx(\mathsf{N}\mathcal{H}+\mathsf{N}^a\mathcal{H}_a
+\mathsf{N}^a_b\mathcal{J}^b_a),
\end{equation}
and since by construction we have that $\mathcal{J}^a_b\approx 0$, the 
argument therefore proceeds exactly as before, with now 
$\mathsf{N}^a:=(\mathsf{N},\mathsf{N}^a,\mathsf{N}^a_b)$.

The decomposition~\eqref{eq:aitch} of the constraints is far from unique, 
however there is at least a natural identification of $\mathcal{J}^a_b$ with 
the generators of $\mathfrak{gl}(n,\mathbb{R})$.
Here we will assume throughout that $\mathcal{J}^a_b$ are the phase space 
representation of the Lie algebra of frame transformations in $\Delta^a_b$ 
as described in~\eqref{eq:rotat}.
In fact, just as the coordinate frame surface diffeomorphism generators 
$\mathcal{H}_i$ may be deduced purely from the fields in 
question~\cite{Isenberg+Nester:1980}, it is also always possible to construct 
the generators $\mathcal{J}^a_b$ purely from the chosen parameterisation of 
phase space.
The argument is a straightforward extension of the following example.

Since we are dealing with a general linear frame on $\Sigma$ there are two 
densities to take into account: $\sqrt{-\mathrm{g}}$ as well as 
$E:=\det({E_i}^a)$.
If we have chosen the lapse function and shift vector to be of weight zero 
(\textit{i.e.}, an ordinary scalar and vector respectively) then all of the 
constraints must be of weight one in each, in particular 
$\mathcal{J}^a_b:=E\sqrt{-\mathrm{g}}J^a_b$ for $J^a_b$ an ordinary tensor.
Consider the simple case of a hypersurface vector field 
$\mathcal{Q}^a:=\sqrt{-\mathrm{g}}Q^a$ and its conjugate 
$\mathcal{P}_a:=EP_a$ that appear as coordinates in phase space.
It is straightforward to see that 
$\mathcal{J}^a_b=\mathcal{Q}^a\mathcal{P}_b
+\delta^a_b\mathcal{Q}^c\mathcal{P}_c$ properly generates frame rotations on 
the $(\mathcal{Q}^a,\mathcal{P}_a)$ sector of phase space (note that effect 
of a frame rotation on the densities is nontrivial) without affecting any 
other canonical variable.
(In fact, the action of $\mathcal{J}^a_b$ is only identical to that of 
$\Delta^a_b$ up to a sign, since $\Delta^a_b$ acts from the left and 
$\mathcal{J}^a_b$ acts from the right via the Poisson bracket; see the 
comments following~\eqref{eq:Fd comms}.)
This is easily generalised to vectors of arbitrary weight by adjusting the 
coefficient of trace term, and higher-order tensors by considering the 
possible $(1,1)$ tensors built from contractions of the coordinate with the 
momenta.

That one may do this is important since it guarantees that the chosen form 
of $\mathcal{J}^a_b$ will satisfy the Lie algebra~\eqref{eq:Gl3} of 
$\mathfrak{gl}(n,\mathbb{R})$ strongly (\textit{i.e.}, on all of phase 
space), and $\mathcal{J}^a_b\approx 0$ are therefore first class constraints, 
closing separately from the diffeomorphism constraints.
In contrast, we will \textit{not} assume any particular form or action the 
diffeomorphism generators (which may clearly mix linearlly with the frame 
rotation generators as, for example 
$\tilde{\mathcal{H}}=\mathcal{H}+\kappa\mathcal{J}^a_a$ and 
$\tilde{\mathcal{H}}_a=\mathcal{H}_a+\lambda\nabla_b[\mathcal{J}]^b_a$).
In particular, we will consider the following two cases:
The first where $\mathcal{H}_a$ acts on the components of tensors as a Lie 
derivative without affecting the vielbeins at all (as in the standard 
coordinate frame approach), and the second where $\mathcal{H}_a$ acts on the 
components of tensors to give a covariant derivative (and therefore the 
action on the surface metric vanishes using metric compatibility) while at 
the same time producing a rotation of the spatial vielbeins.
(In this case there is also a mixing of $\mathcal{H}$ with $\mathcal{J}^a_b$ 
in order to guarantee that the form of the spatial metric is not affected by 
the action of $\mathcal{H}$.)

\section{The Hypersurface Deformation Algebras}

As mentioned in the previous section, we will be considering the deformation 
generators in order to compute the structure functions $\kappa^A_{BC}$.
These generators are defined by~\cite{Hojman+Kuchar+Teitelboim:1976}
\begin{equation}\label{eq:deltas}       
\delta_{\alpha x}:=\delta/\delta\mathsf{e}^\alpha(x),\quad
\delta_{\mathsf{n} x}:=\mathsf{n}^\alpha(x)\delta_{\alpha x},\quad
\delta_{ax}:=\mathsf{e}^\alpha_a(x)\delta_{\alpha x},
\end{equation}
and generate deformations of the image of $\Sigma$ in $\mathbf{M}$.
This may be seen by considering a deformation of $\Sigma$ that is represented 
by an infinitessimal change in the embedding as 
$\mathsf{e}\rightarrow\mathsf{e}+\delta\mathsf{e}$.
A functional of the embedding will then change by 
$F[\mathsf{e}]\rightarrow F[\mathsf{e}+\delta\mathsf{e}]
=F[\mathsf{e}]+V^\alpha\delta_{\alpha x}[F]\rvert_{\mathsf{e}}$, where 
$V^\alpha$ represents the vector along which the surface is deformed.
This is, of course, the basis for the geometry of hyperspace considered 
extensively by Kuchar, and to quote~\cite{Isham+Kuchar:1985a}: ``A reader 
perverse enough to ask for more details i[s] referred 
to~\cite{Kuchar:1976a,Kuchar:1976b}''.
In particular, if we deform $\Sigma$ along a chosen family of embeddings 
represented locally by $\mathsf{N}$ and $\mathsf{N}^a$, then the (coordinate 
frame) deformation operator 
$\int_\Sigma dx\bigl(\mathsf{N}(x)\delta_{\mathsf{n}x}
+\mathsf{N}^i(x)\delta_{ix}\bigr)$ will be identical to the time derivative 
operator $\partial_t$ on all tensors, consistent with the fact that 
$\mathsf{N}$ and $\mathsf{N}^i$ are the normal an tangential projections 
respectively of $\partial_t\in T_t\mathbb{R}$.

Here we will actually have to deal with generalisations of the time 
derivative operator since the components of any tensor are defined with 
respect to a spatial frame, and there is a difference between the partial 
derivative of the components of a tensor and the partial derivative of a 
tensor expanded in the local frame.
Hence we will identify two such oparators, on of which is the partial 
derivative operator that acts on tensor components as 
$\partial_t:T^{ab\cdots}_{mn\cdots}\rightarrow\partial_t
[T^{ab\cdots}_{mn\cdots}]$, and the other is the total derivative operator 
that acts as $\mathrm{d}_t:T^{ab\cdots}_{mn\cdots}\rightarrow
\theta^a\otimes\theta^b\cdots E_m\otimes E_n\cdots
\partial_t[T^{ab\cdots}_{mn\cdots}E_{a^\prime}\otimes 
E_{b^\prime}\cdots\theta^{m^\prime}\otimes\theta^{n^\prime}\cdots]$ 
(\textit{i.e.}, that takes into account the evolution of the frame as well).
Clearly these operators are identical when operating on scalars and the 
tensors themselves (not just the components) and are related, for example 
on a covector field by 
\begin{equation}
\mathrm{d}_t[V_a]
=\bigl[\partial_t[V_b\theta^b]\bigr][E_a]
=\partial_t[V_a]-\partial_t[{E_a}^i]{E_i}^bV_b.
\end{equation}

The linear frame generalization of the variation of $\mathsf{e}_*$ and 
$\mathsf{n}^\alpha$ given 
in~\cite{Teitelboim:1973,Hojman+Kuchar+Teitelboim:1976} may be easily 
computed from~\eqref{eq:emb pf}, yielding
\begin{equation}\label{eq:delta e}
\delta_{\alpha x}[\mathsf{e}^\beta_b(y)]
=\delta^\beta_\alpha E_{by}[\delta(y,x)]
+\mathsf{e}^\beta_a(y)\delta_{\alpha x}[{E_b}^i(y)]{E_i}^a(y)
\end{equation}
for the embedding, and from the variation of~\eqref{eq:normal}
\begin{subequations}\label{eq:delta normal}
\begin{align}
\label{eq:first}
\delta_{\mathsf{n} x}[\mathsf{n}^\mu(y)]
=&\partial_{\mathsf{n}}[\mathsf{n}^\mu(y)]\delta(y,x)
-\mathsf{e}^{\mu b}(y)\bigl(E_{by}[\delta(y,x)]-a_b(y)\delta(y,x)\bigr),\\
\delta_{ax}[\mathsf{n}^\mu(y)]
=&E_{ay}[\mathsf{n}^\mu(y)]\delta(y,x),
\end{align}
\end{subequations}
where $\delta(x,y)$ is defined as a scalar at $x$ and a density at $y$.
These tell us how the pullback and normal change as the surface is deformed 
and depend in general on the chosen family of embeddings via the presence of 
$\mathsf{N}$ in~\eqref{eq:first}.

We begin by choosing the frame on $\Sigma$ such that it is completely 
decoupled from the embedding and geometric structure of $\mathbf{M}$ 
(\textit{i.e.}, $\delta_{\alpha x}[{E_a}^i(y)]=0$).
This does not mean that the vierbein is in any way trivial, just that it is 
chosen without reference to the embedding; we will come back to this later 
on in this section.
Using \eqref{eq:delta e} and \eqref{eq:delta normal}, and the fact that the 
coordinate frame deformation vectors commute 
$[\delta_{\alpha x},\delta_{\beta y}]=0$, we calculate the generalized 
commutator algebra
\begin{subequations}\label{eq:Fd comms}
\begin{align}
\label{eq:perp perp}
[\delta_{\mathsf{n} x},\delta_{\mathsf{n} y}]&=
\mathrm{g}^{ab}(x)E_{ax}[\delta(x,y)]\delta_{bx}
-\mathrm{g}^{ab}(y)E_{ay}[\delta(y,x)]\delta_{by},\\
\label{eq:a perp}
[\delta_{ax},\delta_{\mathsf{n} y}]&=
-E_{ax}[\delta(x,y)]\delta_{\mathsf{n} x},\\
\label{eq:a b}
[\delta_{ax},\delta_{by}]&=
E_{by}[\delta(y,x)]\delta_{ay}
-E_{ax}[\delta(x,y)]\delta_{bx}
-{C_{ab}}^c(y)\delta(y,x)\delta_{cy}.
\end{align}
Due to the scalar and vector quality of $\delta_{\mathsf{n} x}$ and 
$\delta_{ax}$ one finds
\begin{equation}\label{eq:dd}
[\Delta^a_{by},\delta_{\mathsf{n} x}]=0,\quad
[\Delta^a_{bx},\delta_{cy}]
=-\delta^a_c\delta(y,x)\delta_{by},
\end{equation}
\end{subequations}
and~\eqref{eq:Fd comms} combined with~\eqref{eq:Gl3} complete the commutation 
algebra of the set $(\delta_\mathsf{n},\delta_a,\Delta^a_b)$ of surface 
deformations and frame rotations.
(In deriving these, the general rule
\begin{equation}
f(x)\partial_{y^i}\partial_{y^j}\cdots[\delta(y,x)]
=\partial_{y^i}\partial_{y^j}\cdots[f(y)\delta(y,x)],
\end{equation}
is useful, as is the variation of the spacetime point with respect to the 
embedding, as in the case of a scalar 
$\delta_{\alpha x}\bigl[f[\mathsf{e}(y)]\bigr]
=\partial_\alpha[f]\rvert_{\mathsf{e}(y)}\delta(y,x)$.)

There are two ways to see that the structure functions in~\eqref{eq:Fd comms} 
are identical to those appearing in~\eqref{eq:H alg}.
The most straightforward is to note that one would expect the generators of 
hypersurface deformations to have an equivalent action as the generators of 
dynamical evolution (and frame rotation) as given by the Hamiltonian system.
This \textit{is} in fact the case, and was used 
in~\cite{Hojman+Kuchar+Teitelboim:1976} in an identical manner to derive the 
constraint algebra in a coordinate frame.
Therefore we may relate the operators $\delta[\cdot]$ directly to 
$\{\cdot,\mathcal{H}\}$, and merely replace the operators 
$(\delta_{\mathsf{n}},\delta_a,\Delta^a_b)$ in~\eqref{eq:Gl3} 
and~\eqref{eq:Fd comms} with $(\mathcal{H},\mathcal{H}_a,\mathcal{J}^a_b)$ 
(up to a sign since the former act from the left and the latter act from the 
right~\cite{Hojman+Kuchar+Teitelboim:1976}.)
The result of this is that the structure constants $\kappa^A_{BC}$ may be 
read off of~\eqref{eq:Fd comms} directly, and in the coordinate frame limit 
(${E_a}^i(x)=\delta^i_a$) agrees with previous 
results~\cite{Teitelboim:1973,Kuchar:1976a}.
Alternatively one may construct~\eqref{eq:structure functions} directly, 
which results from smearing~\eqref{eq:Fd comms} with $\mathsf{N}^A_1(x)$ and 
$\mathsf{N}^A_2(y)$.
That $\mathsf{N}^A_{1,2}(x)$ should occur outside the commutators is due to 
the fact that by assumption they represent families of embeddings with 
respect to any hypersurface, that is, the components $\mathsf{N}^A_{1,2}(x)$ 
are taken to be the the same on any hypersurface.

That the structure functions should be determined by geometry alone is merely 
a reflection of the fact that the setting itself is purely geometrical; we 
have not commited ourselves to a particular dynamical model, we have stated 
that it should be true \textit{regardless} of the model in question, and in 
fact the method that we will employ here in order to compute these structure 
functions will reflect this.
The original derivation of Teitelboim~\cite{Teitelboim:1973} consisted of a 
direct computation of $\mathsf{N}^A$ from $\mathsf{N}^A_{1,2}$ via Taylor 
expansion, however instead we will follow the more geometric and systematic 
procedure of~\cite{Hojman+Kuchar+Teitelboim:1976} whereby we consider the 
action of the generators of surface deformations directly.
Thus the construction is far more powerful since the structure functions may 
be related to any tensor, and we end up with generators that in fact 
represent the geometrical content of any (covariant) dynamical system 
represented by a Hamiltonain. 

When transferred to the constraints this algebra may be related to the 
Bianchi identities; if the evolution equations are all satisfied, what 
remains are the evolution equations for the 
constraints~\cite{Adler+Bazin+Schiffer:1975}, and $\dot{\mathcal{H}}$ and 
$\dot{\mathcal{H}}_a$ are just combinations of~\eqref{eq:Fd comms}.
This is made more explicit in~\cite{Fischer+Marsden:1979}, where the algebra 
appears as evolution equations for the constraints on the space of 
gravitational degrees of freedom that satisfy the evolution equations but 
not (necessarily) the constraints.
This is sensible since both are consequences of diffeomorphism invariance, 
and thus the constraint algebra may be considered to be the Hamiltonian form 
of the Bianchi identities.

The commutator algebra~\eqref{eq:Fd comms} and~\eqref{eq:Gl3} may also be 
determined explicitly from the action of the generators on various objects.
By assumption, the variations do not affect the frames themselves 
($\delta_{\alpha x}[{E_a}^i]=0$), and their action on tensors above 
$\mathbf{M}$ that have been pulled-back to tensors above $\Sigma$ may be 
determined from the explicit form of the pull-back (\textit{i.e.}, from 
$V_a=\mathsf{e}_a^\alpha V_\alpha$).
The action of the perpendicular generator is perhaps slightly more 
complicated than would be expected, yielding for example on a covector
\begin{equation}\label{eq:perp deltas}
\delta_{\mathsf{n} x}[V_a(y)]=
\partial_{\mathsf{n}y}[V_a(y)]\delta(y,x)
-{C_{\mathsf{n}a}}^b(y)V_b(y)\delta(y,x)
+V_{\mathsf{n}}(y)\bigl(E_{ay}[\delta(y,x)]-a_a(y)\delta(y,x)\bigr),
\end{equation}
which mixes the spatial and perpendicular projections of tensors.
However, if one smears this with respect to the lapse function $\mathsf{N}$, 
one finds the familiar result $\int_{\Sigma}dx\,
\mathsf{N}(x)\delta_{\mathsf{n} x}[V_a(y)]=
\mathsf{N}(y)\bigl(\partial_{\mathsf{n}y}[V_a(y)]
-{C_{\mathsf{n}a}}^b(y)V_b(y)\bigr)$, which is the surface-covariant normal 
derivative operator~\cite{Isenberg+Nester:1980}.
One can explicitly show that the operator $\delta_{\mathsf{n} x}$ does not 
change the $\Sigma$ tensor character of objects (\textit{i.e.}, it is a 
scalar operator, for example if $V_a\rightarrow \lvert M^{-1}\rvert^b_aV_b$, 
then $\delta_{\mathsf{n} x}[V_a]\rightarrow 
\lvert M^{-1}\rvert^b_a\delta_{\mathsf{n} x}[V_b]$) which results in 
(\ref{eq:a perp}).
The tangential generators $\delta_{ax}$ act as, for example
\begin{equation}\label{eq:parallel deltas}
\delta_{ax}[V_b(y)]=E_{ay}[V_b(y)]\delta(y,x)+E_{by}[\delta(y,x)]V_a(y)
-{C_{ab}}^c(y)V_c(y)\delta(y,x),
\end{equation}  
which, when contracted with a vector field and integrated over $\Sigma$, 
yields the Lie derivative defined on $\Sigma$ (\textit{e.g.}, 
$\int_\Sigma dx\,M^a(x)\delta_{ax}[V_b(y)]=\pounds_{\vec{M}}[V]_b(y)$).
Therefore $\delta_{ax}$ is said to generate infinitesimal diffeomorphisms, 
and the algebra \eqref{eq:a b} is that of $\mathit{LDiff}\Sigma$.
The form of the commutators \eqref{eq:a perp} and \eqref{eq:a b} may be 
derived by taking into account the fact that $\delta_{\mathsf{n}x}$ and 
$\delta_{a x}$ are vector and scalar density operators of weight one 
respectively.
With this choice of generators, we naturally find the partial time derivative 
operator $\partial_t=\int_\Sigma dx\,
\bigl(\mathsf{N}(x)\delta_{\mathsf{n}x}+\mathsf{N}^a(x)\delta_{ax}\bigr)$.

The set of generators considered thus far ($\delta_{\mathsf{n} x}$, 
$\delta_{ax}$ and $\Delta^a_{bx}$) is convenient for considering frames that 
have been fixed independently of the foliation, however not for considering 
the opposite case, namely, where the spatial metric has a fixed form.
Explicitly, the action of the generators on the components of the spatial 
metric is
\begin{subequations}
\begin{align}
\delta_{\mathsf{n} x}[\mathrm{g}_{ab}(y)]=&
\partial_{\mathsf{n}y}[\mathrm{g}_{ab}(y)]\delta(y,x)
-{C_{\mathsf{n}a}}^c(y)\mathrm{g}_{cb}(y)\delta(y,x)
-{C_{\mathsf{n}b}}^c(y)\mathrm{g}_{ac}(y)\delta(y,x),\\
\delta_{ax}[\mathrm{g}_{bc}(y)]=&
E_{ay}[\mathrm{g}_{bc}(y)]\delta(y,x)
+E_{by}[\delta(y,x)]\mathrm{g}_{ac}(y)
+E_{cy}[\delta(y,x)]\mathrm{g}_{ba}(y)\nonumber \\
&-{C_{ab}}^d(y)\mathrm{g}_{dc}(y)\delta(y,x)
-{C_{ac}}^d(y)\mathrm{g}_{bd}(y)\delta(y,x),
\end{align}
\end{subequations}
in neither case preserving its form.
This means that if we wanted to specialise to an orthonormal frame on 
$\Sigma$ by taking $\mathrm{g}_{ab}=-\delta_{ab}$, the action of the 
deformation generators (and therefore also the related constraints) would 
not respect this.

In order to deal with this case (which is what is considered 
in~\cite{Henneaux:1983,Charap+Henneaux+Nelson:1988}) we will define the 
following generators which mix the hypersurface deformation generators with 
the generators of frame rotations:
\begin{subequations}\label{eq:prime definitions}
\begin{equation}
\delta^\prime_{\mathsf{n} x}:=\delta_{\mathsf{n} x}-\Delta_x,\quad
\delta^+_{a x}:=\delta_{a x}-\Delta^+_{ax},\quad 
\delta^\prime_{a x}:=\delta_{a x}-\Delta_{ax},
\end{equation}
where 
\begin{equation}
\Delta_x:=k^a_b(x)\Delta^b_{ax},
\end{equation}
and the action of the $\Delta_{ax}$ and $\Delta^+_{ax}$ is defined to be
\begin{align}
\int_\Sigma dx\,f^a(x)\Delta_{ax}=&
-\int_\Sigma dx\nabla_b[f]^a(x)\Delta^b_{ax},\\
\int_\Sigma dx\,f^a(x)\Delta^+_{ax}=&
-\int_\Sigma dx\,\mathrm{g}^{ac}(x)\nabla_{(b}[f]_{c)}(x)\Delta^b_{ax}.
\end{align}
\end{subequations}
(Note that $\Delta^+_{ax}$ is the contribution to $\Delta_{ax}$ from the 
symmetric generators $\Delta^{(ab)}_x$ where 
$\Delta^{ab}_x:=\mathrm{g}^{ac}(x)\Delta^b_{cx}$.
Symmetrization and antisymmetrization on a pair of indices is indicated by 
$(\,)$ and $[\,]$ respectively \textit{e.g.}, 
$T_{[ab]}:=\tfrac{1}{2}(T_{ab}-T_{ba})$.)

In order to compute the algebra based on the set of generators 
$(\delta^\prime_{\mathsf{n}},\delta^\prime_a,\Delta^a_b)$ or 
$(\delta^\prime_{\mathsf{n}},\delta^+_a,\Delta^a_b)$, we will need to 
generate some intermediate results.
In particular, $k_{ab}$ and $\Gamma^a_{bc}$ appear in~\eqref{eq:prime 
definitions} although as it turns out, we will only need to compute their 
normal variations.
That this compution is nontrivial follows from the fact that, unlike most 
of the tensors over $\Sigma$ that we have been considering, $k_{ab}$ and 
$\Gamma^a_{bc}$ are \textit{not} merely surface projections of spacetime 
tensors.
Nevertheless, the dependence on the embedding is given explicitly by the 
spacetime definitions given in~\eqref{eq:k definition} and~\eqref{eq:Gammer} 
respectively.
The required results are:
\begin{subequations}\label{eq:delta k}
\begin{align}
\delta_{\mathsf{n} x}[k_{ab}(y)]=&
\partial_{\mathsf{n}}[k_{ab}(y)]\delta(y,x)
-k_{ac}(y){C_{\mathsf{n}b}}^c(y)\delta(y,x)
-k_{bc}(y){C_{\mathsf{n}a}}^c(y)\delta(y,x) \\
&-\nabla_a[a]_b(y)\delta(y,x)
-a_a(y)a_b(y)\delta(y,x)
+E_{ay}\bigl[E_{by}[\delta(y,x)]\bigr]
-\Gamma^c_{ab}(y)E_{cy}[\delta(y,x)],\nonumber  \\
\delta_{\mathsf{n} x}[\Gamma^a_{bc}(y)]=&
-{}^\Sigma\!{R^a}_{bc\mathsf{n}}(y)\delta(y,x)
-\nabla_b[k]^a_c(y)\delta(y,x)\nonumber \\
&-k^a_b(y)E_{cy}[\delta(y,x)]
-k^a_c(y)E_{by}[\delta(y,x)]
+k_{bc}(y)\mathrm{g}^{ad}(y)E_{dy}[\delta(y,x)].
\end{align}
\end{subequations}
From the scalar and vector nature of $\Delta_x$ and $\Delta_{ax}$ 
respectively, it is easy to compute
\begin{subequations}\label{eq:extra comms}
\begin{equation}
[\Delta_x,\Delta_{y}]=0,\quad
[\Delta^b_{ax},\Delta_{y}]=0,\quad
[\Delta^a_{bx},\Delta_{cy}]=-\delta^a_c\delta(y,x)\Delta_{by},
\end{equation}
and using~\eqref{eq:delta k}, the remaining commutators that are necessary to 
compute the algebra are given by
\begin{align}
[\Delta_{ax},\Delta_{y}]=&
k^b_a(y)\delta(y,x)\Delta_{by}\nonumber \\
&+\bigl(\delta^d_aE_{cy}[\delta(y,x)]+
\Gamma^d_{ca}(y)\delta(y,x)\bigr)
k^b_d(y)\Delta^c_{by}\nonumber \\
&-\bigl(\delta^b_aE_{dy}[\delta(y,x)]+
\Gamma^b_{da}(y)\delta(y,x)\bigr)
k^d_c(y)\Delta^c_{by},\\
[\delta_{\mathsf{n} x},\Delta_{y}]=&
\bigl(\partial_{\mathsf{n}}[k^a_b(y)]\delta(y,x)
+{C_{\mathsf{n}d}}^a(y)k^d_b(y)\delta(y,x)
-{C_{\mathsf{n}b}}^d(y)k^a_d(y)\delta(y,x)\bigr)\Delta^b_{ay}\nonumber \\
&-\mathrm{g}^{ac}(y)\bigl(\nabla_b[a]_c(y)+a_b(y)a_c(y)\bigr)
\delta(y,x)\Delta^b_{ay}\nonumber \\
&+\mathrm{g}^{ac}(y)\bigl(E_{by}\bigl[E_{cy}[\delta(y,x)]\bigr]
-\Gamma^d_{bc}(y)E_{dy}[\delta(y,x)]\bigr)\Delta^b_{ay},\\
[\delta_{\mathsf{n} x},\Delta_{ay}]=&
\bigl({R^b}_{ca\mathsf{n}}(y)\delta(y,x)
+\nabla_c[k]^b_a(y)\delta(y,x)\bigr)\Delta^c_{by}\nonumber \\
&+\bigl(k^b_c(y)E_{ay}[\delta(y,x)]
+k^b_a(y)E_{cy}[\delta(y,x)]
-k_{ac}(y)\mathrm{g}^{bd}(y)E_{dy}[\delta(y,x)]
\bigr)\Delta^c_{by},\\
[\delta_{ax},\Delta_{y}]=&
k^b_a(y)\delta(y,x)\delta_{by}
+\nabla_a[k]^b_c(y)\delta(y,x)\Delta^c_{by}\nonumber \\
&+\bigl(\delta^d_aE_{cy}[\delta(y,x)]
+\Gamma^d_{ca}(y)\delta(y,x)\bigr)k^b_d(y)\Delta^c_{by}\nonumber \\
&-\bigl(\delta^b_aE_{dy}[\delta(y,x)]+
\Gamma^b_{da}(y)\delta(y,x)\bigr)k^d_c(y)\Delta^c_{by}.
\end{align}
\end{subequations}

Using these results, we find the commutator algebra of the set 
($\delta^\prime_{\mathsf{n} x}$, $\delta^\prime_{ax}$, $\Delta^a_{bx}$) to 
be given by~\eqref{eq:Gl3} and
\begin{subequations}\label{eq:prime algebra}
\begin{align}
\label{eq:pperp pperp}
[\delta^\prime_{\mathsf{n} x},\delta^\prime_{\mathsf{n} y}]=&
\mathrm{g}^{ab}(x)E_{ax}[\delta(x,y)]\delta^+_{bx}
-\mathrm{g}^{ab}(y)E_{ay}[\delta(y,x)]\delta^+_{by}, \\
\label{eq:pa pperp}
[\delta^\prime_{ax},\delta^\prime_{\mathsf{n} y}]=&
-E_{ax}[\delta(x,y)]\delta^\prime_{\mathsf{n} x}
-k^b_a(y)\delta(y,x)\delta^\prime_{by}\nonumber\\
&+{R^b}_{ca\mathsf{n}}(y)\delta(y,x)\Delta^c_{by}
+2k_{a[b}(x)E_{c]x}[\delta(x,y)]\Delta^{bc}_{x},\\
\label{eq:pa pb}
[\delta^\prime_{ax},\delta^\prime_{by}]=&
{}^\Sigma\!{R^c}_{dab}(y)\delta(y,x)\Delta^d_{cy},\\
[\Delta^a_{by},\delta^\prime_{\mathsf{n} x}]=&0,\qquad\qquad
[\Delta^a_{bx},\delta^\prime_{cy}]
=-\delta^a_c\delta(y,x)\delta^\prime_{by},
\end{align}
\end{subequations}
where these new generators act on tensors to give, for example
\begin{subequations}\label{eq:ssstensors}
\begin{align}
\delta^\prime_{\mathsf{n} x}[V_a(y)]=&
\partial_{\mathsf{n}y}[V_a(y)]\delta(y,x)
-{C_{\mathsf{n}a}}^b(y)V_b(y)\delta(y,x)
+k_a^b(y)V_b(y)\delta(y,x)\nonumber \\
&\quad 
+V_{\mathsf{n}}(y)\bigl(E_{ay}[\delta(y,x)]-a_a(y)\delta(y,x)\bigr),\\
\delta^\prime_{a x}[V_b(y)]=&\nabla_{a}[V]_b(y)\delta(y,x),
\end{align}
\end{subequations}
and rotates the vielbeins through
\begin{subequations}\label{eq:primed viels}
\begin{align}
\delta^\prime_{\mathsf{n} x}[{E_b}^j(y)]=&
k^a_b(y){E_a}^j(y)\delta(y,x),\\
\delta^\prime_{ax}[{E_b}^j(y)]=&
-\bigl(\delta^c_aE_{by}[\delta(y,x)]
+\Gamma^c_{ba}(y)\delta(y,x)\bigr){E_c}^j(y).
\end{align}
\end{subequations}
The commutator~\eqref{eq:pa pb} was derived by noting that $\delta^\prime_a$ 
acts on tensors as a covariant derivative, the commutator of which results in 
the curvature operator; this is why only the normal variations of $k_{ab}$ 
and $\Gamma^a_{bc}$ were required.

In contrast to the unprimed generators, the action of this new set of 
generators on the components of the spatial metric is
\begin{subequations}\label{eq:dprime g}
\begin{align}
\delta^\prime_{\mathsf{n} x}[\mathrm{g}_{ab}(y)]=&
\partial_{\mathsf{n}y}[\mathrm{g}_{ab}(y)]\delta(y,x)
-{C_{\mathsf{n}a}}^c(y)\mathrm{g}_{cb}(y)\delta(y,x)
-{C_{\mathsf{n}b}}^c(y)\mathrm{g}_{ac}(y)\delta(y,x)\nonumber \\
&+k^c_a(y)\mathrm{g}_{cb}(y)\delta(y,x)
+k^c_b(y)\mathrm{g}_{ac}(y)\delta(y,x), \\
\delta^\prime_{ax}[\mathrm{g}_{bc}(y)]=&
\nabla_a[\mathrm{g}]_{bc}(y)\delta(y,x),
\end{align}
\end{subequations}
both of which vanish due to \eqref{eq:k definition} and the spatial metric 
compatibility conditions respectively.
It is this set of deformation generators that represent the total time 
derivative operator $\mathrm{d}_t=\int_\Sigma dx\,
\bigl(\mathsf{N}(x)\delta^\prime_{\mathsf{n}x}
+\mathsf{N}^a(x)\delta^\prime_{ax}\bigr)$ and correspond to the mixing of the 
original diffeomorphism constraints by
\begin{equation}
\mathcal{H}^\prime:=\mathcal{H}-k^a_b\mathcal{J}^b_a,\quad
\mathcal{H}^\prime_a:=\mathcal{H}-\nabla_b\mathcal{J}^b_a.
\end{equation}
Due to~\eqref{eq:dprime g}, these constraints will not alter the form of the 
spatial metric, and the reduction to orthonormal frames on $\Sigma$ is a 
simple matter of choosing $\mathrm{g}_{ab}=-\delta_{ab}$ and restricting 
$\Delta^a_b$ (or $\mathcal{J}^a_b$) to correspond to generators of 
$\mathfrak{so}(n)$.

Furthermore, defining the coordinate components 
$\delta^\prime_{ix}:={E_i}^a(x)\delta^\prime_{ax}$ and using 
\eqref{eq:primed viels}, the constraint algebra related to the coordinate 
generators ($\delta^\prime_{\mathsf{n}},\delta^\prime_{i},\Delta^a_b$) is 
identically the time gauge result given in~\cite{Charap+Henneaux+Nelson:1988}:
\begin{subequations}
\begin{align}
[\delta^\prime_{\mathsf{n} x},\delta^\prime_{\mathsf{n} y}]=&
\mathrm{g}^{ij}(x)\partial_{x^i}[\delta(x,y)]\delta^+_{jx}
-\mathrm{g}^{ij}(y)\partial_{y^i}[\delta(y,x)]\delta^+_{jy}, \\
[\delta^\prime_{ix},\delta^\prime_{\mathsf{n} y}]=&
-\partial_{x^i}[\delta(x,y)]\delta^\prime_{\mathsf{n} x}
+{R^b}_{ci\mathsf{n}}(y)\delta(y,x)\Delta^c_{by}
+2{E_i}^a(x)k_{a[b}(x)E_{c]x}[\delta(x,y)]\Delta^{bc}_x,\\
[\delta^\prime_{ix},\delta^\prime_{jy}]=&
\partial_{y^j}[\delta(y,x)]\delta^\prime_{iy}
-\partial_{x^i}[\delta(x,y)]\delta^\prime_{jx}
+{}^\Sigma\!{R^c}_{dij}(y)\delta(y,x)\Delta^d_{cy},\\
[\Delta^a_{by},\delta_{\mathsf{n} x}]=&0,\qquad\qquad
[\Delta^a_{bx},\delta_{iy}]=0.
\end{align}
\end{subequations}
If one were to consider instead the set ($\delta^\prime_{\mathsf{n} x}$, 
$\delta^+_{ax}$, $\Delta^a_{bx}$), then \eqref{eq:pa pperp} and 
\eqref{eq:pa pb} would be replaced by the following:
\begin{subequations}\label{eq:plus algebra}
\begin{align}
\int_\Sigma dx\,dy\,A^a(x)B(y)
[\delta^+_{ax},\delta^\prime_{\mathsf{n} y}]=&
\int_\Sigma dx\,\bigl(
-A^a\nabla_a[B]\delta^\prime_{\mathsf{n} x}
-BA^ak^b_a\delta^+_{bx}\nonumber \\
&\qquad +2A^ak_{ab}\nabla_c[B]\Delta^{[bc]}_x
+2Bk^c_a\nabla_{[c}[A]_{b]}\Delta^{[ab]}_x\bigr),\\
\int_\Sigma dx\,dy\,A^a(x)B^b(y)
[\delta^+_{ax},\delta^+_{by}]=&
\int_\Sigma dx\,\mathrm{g}^{ab}\bigl(
B^c\nabla_{[c}[A]_{b]}
+A^c\nabla_{[c}[B]_{b]}
\bigr)\delta^+_{ax}\nonumber \\
&-2\int_\Sigma dx\,\mathrm{g}^{cd}
\nabla_{(a}[A]_{c)}
\nabla_{(b}[B]_{d)}\Delta^{[ab]}_x,
\end{align}
\end{subequations}
which have been smeared over an appropriate choice of tensor field for ease 
of display.
The coordinate components related to \eqref{eq:plus algebra} correspond to 
the those initially derived 
in~\cite[Equation (2.6)]{Charap+Henneaux+Nelson:1988}, before the additional 
frame rotation has been performed.

Instead of requiring that the frame be completely independent of the 
embedding, it may be constrained so that the action of the deformation 
generators on the components of the spatial metric is trivial: 
$\delta_{\alpha x}[\mathrm{g}_{ab}(y)]=0$, yielding the condition
\begin{equation}\label{eq:delta Lorentz}
\delta_{\alpha x}[{E_{(a}}^i(y)]E_{ib)}(y)
=-\tfrac{1}{2}\delta_{\alpha x}[\mathrm{g}_{\mu\nu}(y)]
\mathsf{e}^\mu_a(y)\mathsf{e}^\nu_b(y)
-\mathsf{e}_{\alpha(a}(y)E_{b)y}[\delta(y,x)].
\end{equation}
Choosing the vielbein to satisfy
\begin{subequations}
\begin{align}\label{eq:delta Vier}
\delta_{\mathsf{n} x}[{E_a}^i(y)]{E_i}^b(y)
=&k^b_a(y)\delta(y,x),\\
\label{eq:D symm}
\delta_{ax}[{E_b}^i(y)]{E_i}^c(y)
=&-\bigl(\delta^c_aE_{by}[\delta(y,x)]
+\Gamma^c_{ba}(y)\delta(y,x)\bigr),
\end{align}
\end{subequations}
we find that the derived generators and algebra are equivalent to that of 
($\delta^\prime_{\mathsf{n} x},\delta^\prime_{ax},\Delta^a_{bx}$) considered 
above, and removing the antisymmetric part of the right hand side of 
(\ref{eq:D symm}), one recovers the generator $\delta^+_{ax}$.
Thus we have two ultimately identical ways of approaching the problem; the 
first by looking for an equivalent form of the generators that preserve the 
form of the spatial metric, and the second is to constrain the vierbein so 
that the same condition holds.

Clearly there are other representations of the constraint algebra that 
correspond to different ways of mixing the diffeomorphism constraints 
with the generators of frame rotations, however the two cases that we have 
dealt with here have the most direct physical interpretation.
Note that it is also possible to determine how the algebra is altered by a 
(partial) fixing of the remaining $\mathrm{SO}(n)$ invariance.
This would show up in an additional condition on the frame that would have 
to be maintained under the action of the generators, leading to further 
alterations of the constraint algebra.

\section*{Conclusions}
\label{sect:Concl}

What we have developed herein is, in fact, a general formalism for 
determining how a choice of frame affects the diffeomorphism constraint 
algebra of a theory.
The two important limits, namely, that of coordinate frame diffeomorphism 
generators and that of orthonormal frame generators, are reached through a 
mixing of the diffeomorphism constraints with the generators of the 
generalised frame rotations.
In both of these limits, the standard results (of~\cite{Teitelboim:1973} 
and~\cite{Charap+Henneaux+Nelson:1988} respectively) are recovered, however 
since the analysis has not been restricted to a particular model, represents 
the generalisation of the latter results to the algebra of any covariant 
model written in a tetrad frame.

The conclusion drawn from these results is that the diffeomorphism constraint 
algebra is purely a geometrical relation \textit{in any frame}.
Once one has chosen the coordinate system and frames of reference in which a 
particular system is to be described, the constraint algebra of the 
generators of the diffeomorphism algebra is fixed independently of the model.
In any consistent quantisation of the model, the operators that play the 
role of these generators must faithfully represent the algebra on the Hilbert 
space in question.
Since classical (and presumably quantum) General Relativity should relate 
observations made in different frames of reference, one would like any 
potential quantisation (of General Relativity or of quantum fields on a 
curved background) to reproduce this algebra not just in a particular choice 
of frame, but for any choice of frame.

It is also interesting to note that any covariant combination of the 
generators has vanishing commutator with itself providing there is no 
dependence on the embedding other than that due to the spacetime point.
In particular, taking $\delta_V:=V^\alpha\delta_\alpha$ (for some future 
pointing vector field $V^\alpha$, guaranteeing that $\delta_V$ generates 
deformations of the embedded surface forwards with respect to the foliation), 
one finds that $[\delta_{Vx},\delta_{Vy}]=0$ and the rest of the algebra 
in \eqref{eq:Fd comms} remains the same.
The resulting algebra does not depend on canonical coordinates and is 
therefore a true Lie algebra.
Another special case of this which is more relevant to the case of vacuum 
general relativity is the combination 
$\delta_\mathrm{g}:=\mathrm{g}^{\alpha\beta}\delta_\alpha\delta_\beta
=\delta_{\mathsf{n}}^2+\mathrm{g}^{ab}\delta_a\delta_b$ recently discussed 
in~\cite{Kuchar+Romano:1995}.
It would be interesting to determine whether the more general combinations 
discussed in~\cite{Markopoulou:1996} could also be considered as covariant 
combinations of the coordinate frame constraints, and indeed, what the action 
of the resulting generators would be.

\section*{Acknowledgements}

The author would like to thank the Natural Sciences and Engineering Research 
Council of Canada and the University of Toronto for financial support, as 
well as E. Prugove\v{c}ki and F. Hehl for related information, 
J. L\'{e}gar\'{e} for comments on the manuscript, and a sympathetic referee 
for helping this manuscript make sense.

\newpage

\bibliographystyle{amsplain}

\providecommand{\bysame}{\leavevmode\hbox to3em{\hrulefill}\thinspace}

\end{document}